\documentclass[aps,prc,superscriptaddress,twoside,twocolumn,10pt,floatfix]{revtex4-1}
\usepackage{graphicx}
\usepackage{amsmath}
\usepackage{amssymb}
\usepackage{bm}
\usepackage{braket}
\usepackage{color}
\usepackage{hyperref}
\usepackage{mathrsfs}

\hypersetup{bookmarksnumbered,%
  colorlinks,%
	linkcolor=blue,%
  citecolor=blue,%
  urlcolor=blue,
	plainpages=false,%
	pdfstartview=FitH}

\newcommand{\nc}{\newcommand}           % new command
\nc{\figwidth}{0.45\textwidth}                    % for final submission
\begin{document}
\title{Evolution of clustering structure through the momentum distributions\\
in $^{8-10}$Be isotopes}

\author{Songjie Li} \email[]{songjie.li@smail.nju.edu.cn}
\affiliation{School of Physics, Nanjing University, Nanjing 210093, China}

\author{Takayuki Myo} \email[]{takayuki.myo@oit.ac.jp}
\affiliation {General Education, Faculty of Engineering, Osaka Institute of
Technology, Osaka, Osaka 535-8585, Japan} \affiliation {Research Center for
Nuclear Physics (RCNP), Osaka University, Osaka 567-0047, Japan}

\author{Qing Zhao} \email[]{zhao@nucl.sci.hokudai.ac.jp}
\affiliation{Department of Physics, Hokkaido University, Sapporo 060-0810, Japan}

\author{Hiroshi Toki}
\affiliation {Research Center for Nuclear Physics (RCNP), Osaka University,
Osaka 567-0047, Japan}

\author{Hisashi Horiuchi}
\affiliation {Research Center for Nuclear Physics (RCNP), Osaka University,
Osaka 567-0047, Japan}

\author{Chang Xu} \affiliation{School of Physics,
Nanjing University, Nanjing 210093, China}

\author{Jian Liu} \affiliation{College of Science, China University of
Petroleum East China, Qingdao 266580, China}

\author{Mengjiao Lyu} \email[]{corresponding author: mengjiao@rcnp.osaka-u.ac.jp}
\affiliation{Research Center for Nuclear Physics (RCNP), Osaka University, Osaka
567-0047, Japan}
    
\author{Zhongzhou Ren} \email[]{corresponding author: zren@tongji.edu.cn}
\affiliation{School of Physics Science and Engineering, Tongji University,
Shanghai 200092, China}

\date{\today}
\begin{abstract}
We investigate the evolution of clustering structure through the momentum
distributions in the $^{8-10}$Be isotopes. The nucleon dynamics within the
inter-cluster antisymmetrization are discussed via the momentum distribution
of a Brink type $\alpha$-$\alpha$ wave function. For the state with a small
$\alpha$-$\alpha$ distance, we observe a significant depression with a dip
structure at zero-momentum and an enhanced tail at relatively higher momentum
region. In addition, we find the ``cluster structure'' in the intrinsic frame
of momentum space, which is complementary to its significant $\alpha$-cluster
dissolution in the coordinate space because of the strong antisymmetrization.
For the physical $^{8-10}$Be isotopes, the Tohsaki-Horiuchi-Schuck-R{\"o}pke
(THSR) wave functions are adopted. The evolution from the dilute clustering
state to the compact one is demonstrated by a successive depression at the
zero-momentum of nucleon distribution for the two $\alpha$-clusters
within $^{8-10}$Be isotopes. For the compact $^{10}$Be nucleus, the
momentum distribution of all nucleons shows significant depression at
zero-momentum with a dip structure, which is found to be contributed by both
the inter-cluster antisymmetrization and the $p$-orbit occupation of the
valence neutrons. This study proposes a new window for the investigations of
the $\alpha$-clustering effects via the low-momentum components of nuclei,
which is expected to be extended to the heavier nuclear clustering states.
\end{abstract}

\maketitle

\section{Introduction}
In atomic nuclei, strongly correlated nucleons compose spatially localized
subsystems, namely the nuclear clusters \cite{horiuchi86}. The relative motion
between the $\alpha$-clusters is the fundamental mode of dynamics in various
nuclear systems, such as $^8$Be, $^{12}$C, $^{16}$O, and $^{20}$Ne. These
clustering states have been studied by using different theoretical models, as
reviewed in Refs.~\cite{zhou19, freer18, ren18, tohsaki17, schuck16, kimura16,
funaki15, ito14, enyo12, horiuchi12}. 

The Beryllium isotopes are well known for their clustering structures in the
ground states, as discussed in previous theoretical studies \cite{Eny95,
Eny99, Ita00, Ito04, myo14, myo15, fun02, Lyu15, Lyu16}. In these works, many
interesting physical phenomena, including the formation of two $\alpha$
clusters \cite{Eny95,Eny99}, occupation of nuclear molecular orbits by the
valence neutrons \cite{Ita00}, di-cluster configurations \cite{ito14},
contribution from tensor force \cite{myo14, myo15}, and nonlocalized dynamics
of the two $\alpha$-clusters \cite{Lyu16}, have been studied. It is found that
the low lying spectrum of the $^{8-10}$Be isotopes can be well described by
the Tohsaki-Horiuchi-Schuck-R{\"o}pke (THSR) wave functions \cite{fun02,
Lyu15, Lyu16}, which were originally proposed for the description of
$\alpha$-condensate states \cite{tohsaki01} and have been applied to many
clustering phenomena in nuclei \cite{fun02, Lyu15, Lyu16, zhao18, zhao19,
zhou13, zhou14, zhou16}. 

In previous works, the physical properties of cluster states in nuclei, such
as the energy spectrum and the charge radii, have been reproduced by
theoretical calculations. In addition, for probing the clustering effects,
observables in various cluster-involved nuclear reactions have been
investigated, such as the monopole transition strengths \cite{yang14, chi15},
the proton induced $\alpha$-knockout cross section \cite{lyu18, lyu19}, and
the $\alpha$-emission cross sections in the fusion-evaporation reactions
\cite{wang19}. In this work, we propose to probe the evolution of the
$\alpha$-clustering structure through the momentum distributions in the
$^{8-10}$Be isotopes. In general, the momentum distributions could be
extracted from the electron scattering reactions \cite{ciofi96}. 

In recent decades, the electron scattering observables have been adopted to
study the high-momentum components \cite{ciofi96, hen14, ciofi15, hen17} or
the deformation \cite{wang20, liu19} of nuclei. It is found that the tensor
and short-range components of inter-nucleon correlations, induced by the
nuclear force, dominate at the momentum regions at about 2 fm$^{-1}$ and 4
fm$^{-1}$, respectively \cite{lyu19b}. Comparing to the inter-nucleon
correlations, the $\alpha$-correlation dominates at much lower momentum
region, and hence we may connect the $\alpha$-cluster dynamics to the nucleon
momentum distributions below the Fermi momentum of $1.4$ fm$^{-1}$.
Especially, due to the antisymmetrization effect between $\alpha$-clusters, we
expect that the momentum distribution would be depressed or enhanced in the
compact systems where $\alpha$-clusters have strong spatial overlap with each
other, by analogy with the inter-nucleon contacts in the correlated $NN$ pairs
\cite{zhao19b}. Through the momentum distributions in the $^{8-10}$Be
isotopes, we expect to reveal the evolution of the corresponding clustering
structure, from the dilute gas-like state to the compact one with cluster
dissolution. 

For the theoretical description of Beryllium isotopes, we adopt the THSR wave
function developed in our previous works \cite{Lyu15, Lyu16}. In
Ref.~\cite{hirai11}, the momentum distribution of $^{9}$Be was discussed
through the Antisymmetrized Molecular Dynamics (AMD) wave function, but the
center-of-mass component is not treated in the formulation. In this work, the
momentum distributions are predicted by using the analytical derivation
formulated recently in Ref.~\cite{lyu19b}, with subtraction of center-of-mass
motion. 

This paper is organized as follows. In Sec.~\ref{sec:formulation}, we
introduce the formulation of the clustering wave functions and nucleon
momentum distributions for the $^{8-10}$Be isotopes. We note that the one-body
momentum distributions of nucleons are investigated in this work, which are
affected by the inter-nucleon and clustering correlations. In
Sec.~\ref{subsec:brink}, the relation between the $\alpha$-$\alpha$ distance
and the nucleon momentum distribution is discussed through Brink type wave
functions of two $\alpha$-cluster system. In Sec.~\ref{subsec:res-thsr}, we
discuss the nucleon momentum distribution for the $^{8-10}$Be isotopes
predicted by using the physical THSR wave functions, and present their
relation with the evolution of $\alpha$-clustering structure. The last
Sec.~\ref{sec:conclusion} contains the conclusion.

\section{Formulations}
\label{sec:formulation}
We introduce briefly the formulations for the clustering wave functions of the
$^{8-10}$Be isotopes and the corresponding nucleon momentum distributions.
Detailed introductions can be found in Refs.~\cite{Lyu15, Lyu16, lyu19} for
the wave functions and in Ref.~\cite{lyu19b} for the nucleon momentum
distributions.

\subsection{Wave functions of $^{8-10}$Be isotopes}
\label{subsec:wave functions}
We start by writing the Brink wave function of the $^{8}$Be nucleus as
\cite{Bri66}
\begin{equation}\label{eq:brink}
	\begin{aligned}
		\Psi^{\textup{Brink}}(^{8}\text{Be}) = \sqrt{\frac{4!\cdot4!}{8!}} \mathcal{A}
		\{\phi_{\alpha_1}(\bm{R}_1)\phi_{\alpha_2}(\bm{R}_2)\},
	\end{aligned}
\end{equation}
where $\mathcal{A}$ is the antisymmetrizer and the wave function of each alpha
cluster $\phi_\alpha(\bm{R})$ is defined as
\begin{equation}\label{eq:alpha}
	\phi_\alpha(\bm{R}) = \frac{1}{\sqrt{4!}} \mathcal{A}
	\{\phi_{1}(\bm{r}_1,\bm{R})\dots\phi_{4}(\bm{r}_4,\bm{R})\}.
\end{equation}
The single nucleon wave functions $\phi(\bm{r}, \bm{R})$ with the position
$\bm{r}$ are defined as the Gaussian wave packets 
\begin{equation}\label{eq:wf-nucleon}
  \phi(\bm{r},\bm{R}) 
  = (2\nu/\pi)^{3/4} e^{-\nu(\bm{r}-\bm{R})^2}\chi_{\sigma,\tau}
\end{equation}
with centroids $\bm{R}$ and width parameter $\nu=0.27$ fm$^{-2}$ to reproduce
the binding energy of $\alpha$-cluster. The component $\chi_{\sigma,\tau}$ is
for the spin $\sigma$ and isospin $\tau$ of each nucleon.

The Brink wave function in Eq.~(\ref{eq:brink}) describes the localized
configuration of two $\alpha$-clusters in the $^{8}$Be nucleus. In physical
nuclei, it is known that the $\alpha$-clusters perform nonlocalized motion,
which is confined by the Gaussian container in the THSR wave function
\cite{zhou14}
\begin{equation}\label{eq:thsr-8be}
	\begin{aligned}
		\Psi^{\textup{THSR}}(^{8}\text{Be})
    & = \int d\bm{R}_1 d\bm{R}_{2}
    \mathcal{G}(\bm{R}_{1},\bm{\beta}_{\alpha})
    \mathcal{G}(\bm{R}_{2},\bm{\beta}_{\alpha})
		\\
    &\qquad\,\times \mathcal{A}\{
      \phi_{\alpha,1}(\bm{R}_1)\phi_{\alpha,2}(\bm{R}_2)
    \},
	\end{aligned}
\end{equation}
where the container function $\mathcal{G}$ is the deformed Gaussian
\begin{equation}\label{eq:Gauss}
  \mathcal{G}(\bm{R},\bm{\beta})=
  \exp \left(-\frac{R_{x}^{2}
    +R_{y}^{2}}{\beta_{xy}^{2}}
  -\frac{R_{z}^{2}}{\beta_{z}^{2}}
  \right).
\end{equation}
Here, the size of the Gaussian container is determined by the width parameters
${\beta}_{xy}$ and ${\beta}_{z}$ in each direction. For the $^{9,10}$Be
isotopes, we introduce additional valence neutrons into the THSR wave function
as
\begin{align}
\Psi^{\textup{THSR}}(^{9}\text{Be})
  & = \int d\bm{R}_1 d\bm{R}_{2}
    \mathcal{G}(\bm{R}_{1},\bm{\beta}_{\alpha})
    \mathcal{G}(\bm{R}_{2},\bm{\beta}_{\alpha})
    \nonumber\\
  &\qquad\,\times \mathcal{A}\{
      \phi_{\alpha,1}(\bm{R}_1)\phi_{\alpha,2}(\bm{R}_2)
      \phi_{9}^{\pi}
    \},\label{eq:9be}\\
\Psi^{\textup{THSR}}(^{10}\text{Be})
  & = \int d\bm{R}_1 d\bm{R}_{2}
    \mathcal{G}(\bm{R}_{1},\bm{\beta}_{\alpha})
    \mathcal{G}(\bm{R}_{2},\bm{\beta}_{\alpha})
    \nonumber\\
  &\qquad\,\times \mathcal{A}\{
      \phi_{\alpha,1}(\bm{R}_1)\phi_{\alpha,2}(\bm{R}_2)
      \phi_{9}^{\pi}\phi_{10}^{\pi}
    \}.\label{eq:10be}
\end{align}
where $\phi_{9,10}^{\pi}$ are the wave functions of valence neutrons occupying
$\pi$-orbits, which are formulated as
\begin{align}
\phi^{\pi}_{9}(\bm{r})
  &=\int d\bm{R}_{9}
    \mathcal{G}(\bm{R}_{9},\bm{\beta}_{n})
    e^{ i\phi_{\bm{R}_{9}}}
    \phi_{n\uparrow}(\bm{r}_{9},\bm{R}_{9}),
    \label{eq:pi-9be}\\
\phi^{\pi}_{10}(\bm{r})
  &=\int d\bm{R}_{10}
    \mathcal{G}(\bm{R}_{10},\bm{\beta}_{n})
    e^{- i\phi_{\bm{R}_{10}}}
    \phi_{n\downarrow}(\bm{r}_{10},\bm{R}_{10}).
    \label{eq:pi-10be}
\end{align}
Here $\phi_{n\uparrow}$ and $\phi_{n\uparrow}$ are neutron wave functions
defined in Eq.~(\ref{eq:wf-nucleon}) with spin up and down, respectively, and
$\phi_{\bm{R}}$ is the azimuthal angle of the neutron generator coordinate
$\bm{R}$. The exponential factors $e^{ \pm i\phi_{\bm{R}}}$ is introduced to
reproduce the negative parity of the $\pi$-orbit \cite{Lyu15}. More details
for the formulation of Eqs.~(\ref{eq:pi-9be}) and (\ref{eq:pi-10be}) can be
found in Ref.~\cite{Lyu15}. The deformation parameters $\bm{\beta}$ are
optimized by variational calculation for each isotope \cite{Lyu15,Lyu16}.

\subsection{Momentum distribution of the wave functions}
The nucleon momentum distribution operator $\hat{n}(\bm{k})$ for mass number
$A$ is defined in the momentum space as
\begin{equation}
\hat{n}(\bm{k})\equiv
  \sum_{i=1}^{A}\delta\left(\bm{k}_{i}-\bm{k}_{\textrm{G}}-\bm{k}\right),
\end{equation}  
where $\bm{k}_{i}$ is the single-nucleon momentum and $\bm{k}_{G}$ is the
center-of-mass momentum
\begin{equation}
  \bm{k}_{\textrm{G}}=\sum_{i=1}^{A}\bm{k}_{i}.    
\end{equation}
For the AMD wave functions
\cite{enyo12}
\begin{equation}\label{eq:amd}
\Psi^{\text{AMD}}=
  \mathcal{A}\left\{
    \prod_{i=1}^{A} \phi_{i}(\bm{r}_{i},\bm{R}_{i})
  \right\},
\end{equation}
the nucleon momentum distribution with correct treatment for the
center-of-mass motion is written as \cite{lyu19b}
\begin{equation}\label{eq:amd-momentum}
\begin{aligned}
n(\bm{k})
&=\braket{\Psi^{\text{AMD}}|\hat{n}(\bm{k})|\Psi^{\text{AMD}}}\\
&=\sum_{i=1}^{A}n_{i}(\bm{k}),
\end{aligned}
\end{equation}
where $n_{i}(\bm{k})$ is the momentum distribution of each nucleon
\begin{equation}\label{eq:amd-momentum2}
\begin{aligned}
&n_{i}(\bm{k})=
  \left(\frac{1}{2\pi\nu\epsilon}\right)^{3/2}\\
&\qquad\quad\times 
  \sum_{j=1}^{A}\exp\left[
      -\frac{1}{2\nu\epsilon}\left(
          \bm{k}-i\nu(\bm{R}^*_{i}-\bm{R}_{j})
        \right)^{2}
  \right]B_{ij}B^{-1}_{ji},
\end{aligned}
\end{equation}
$\epsilon=(A-1)/A$, and $B_{ij}=\braket{\phi_{i}|\phi_{j}}$ is the overlap
matrix of single nucleon states. We note that in this formulation, the
center-of-mass motion is correctly treated, as discussed in
Ref.~\cite{lyu19b}. The momentum distribution $n(\bm{k})$ in
Eq.~(\ref{eq:amd-momentum}) satisfies the normalization condition
\begin{equation}
  \int d\bm{k}\, n(\bm{k})=A.
\end{equation}
Similarly, we define the proton momentum distribution as
\begin{equation}
n^{P}(\bm{k})=
  \sum_{i\in p}n_{i}(\bm{k}),
\end{equation}
where subscript $i$ denotes all the protons. For the superposed AMD wave
function $\ket{\Psi} = \sum_{a}c_{a} \ket{\Psi_{a}}$, the corresponding
nucleon momentum distribution is given as
\begin{equation}\label{eq:amd-sum}
\begin{aligned}
n(\bm{k})=&
  \frac{1}
    {\braket{\Psi|\Psi}}
  \sum_{a,b}
    c_{a}^{*}c_{b}\braket{\Psi_{a}|\widehat{n}|\Psi_{b}},\\
\end{aligned}
\end{equation}
where $c_{a}$ and $c_{b}$ are superposition coefficients.

Mathematically, the Brink wave function in Eq.~(\ref{eq:brink}) can be written
as a special case of the AMD wave function
\begin{equation}\label{eq:brink-amd}
\begin{aligned}
&\Psi^{\textup{Brink}}(^{8}\text{Be}) \\
&\quad= \sqrt{\frac{1}{8!}} \mathcal{A}
  \{\phi_{1}(\bm{r}_{1},\bm{R}_1)\dots\phi_{4}(\bm{r}_{4},\bm{R}_1)\\
&\qquad\qquad\quad\ 
  \times\phi_{5}(\bm{r}_{5},\bm{R}_2)\dots\phi_{8}(\bm{r}_{8},\bm{R}_2)\},
\end{aligned}
\end{equation}
which has the same format as Eq.~(\ref{eq:amd}). In addition, the THSR wave
functions in Eqs.~(\ref{eq:thsr-8be}), (\ref{eq:9be}), and (\ref{eq:10be}) are
mathematically equivalent to the superposed AMD wave functions. As an example,
we write the case for $^{9}$Be as
\begin{align}\label{eq:9be-amd}
&\Psi^{\textup{THSR}}(^{9}\text{Be})=\nonumber\\
&\quad\int d\bm{R}_1 d\bm{R}_{2}
  \mathcal{G}(\bm{R}_{1},\bm{\beta}_{\alpha})
  \mathcal{G}(\bm{R}_{2},\bm{\beta}_{\alpha})
  \nonumber\times\\
&\qquad\int d\bm{R}_{9}
  \mathcal{G}(\bm{R}_{9},\bm{\beta}_{n})
  e^{ i\phi_{\bm{R}_{9}}}
  \nonumber\times\\
&\qquad\quad\ 
  \mathcal{A}\{
  \phi_{1}(\bm{r}_{1},\bm{R}_1)\dots\phi_{8}(\bm{r}_{8},\bm{R}_2)
    \phi_{n\uparrow}(\bm{r}_{9},\bm{R}_{9})
  \}.
\end{align}
Hence, the momentum distributions of the Brink or THSR wave functions in this
work can be calculated using the analytical formulations in
Eqs.~(\ref{eq:amd-momentum}) and (\ref{eq:amd-sum}).

\section{Results}
\subsection{Nucleon momentum distribution of the $\alpha$-$\alpha$ system}
\label{subsec:brink}
\begin{figure}[h]
  \centering
	\includegraphics[width=0.5 \textwidth]{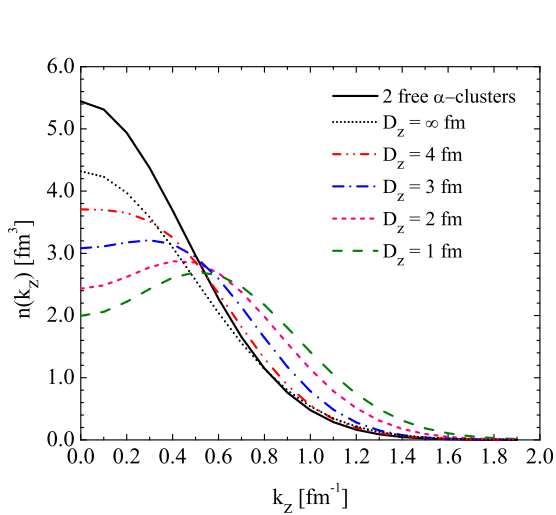}
	\caption{The intrinsic momentum distributions of the $^{8}$Be nucleus
  described by the Brink wave function along the $z$-axis. ``$D_z$'' is the
  relative distance between $\alpha$ clusters in $z$-axis. ``2 free
  $\alpha$-clusters'' denotes the nucleon momentum distribution of two free
  $\alpha$-clusters.}
	\label{fig:brink-curve}
\end{figure}

We first show the nucleon momentum distribution of two free $\alpha$-clusters
as the solid curve in Fig.~\ref{fig:brink-curve}. It is found that the
momentum of free $\alpha$-clusters distributes in a Gaussian form, which is
the Fourier transformation of the $\alpha$-cluster wave function in
Eq.~(\ref{eq:alpha}). We note that this Gaussian distribution is only valid in
low momentum region less than the Fermi momentum $k_{\rm{F}}=1.4$ fm$^{-1}$.
As predicted in Ref.~\cite{lyu19b}, the high-momentum region is dominated by
the tensor and short-range correlations around $k\approx 2$ fm$^{-1}$ and
$k\approx 4$ fm$^{-1}$, respectively. In this work, we focus on the clustering
correlation and choose the effective Volkov $NN$ interaction which does not
include the tensor component or the short-range repulsion, and we limit our
discussion on the momentum with up to $k_{\rm{F}}$ to avoid the effect from the
high-momentum component. In addition, the G3RS term is adopted for the
spin-orbit interaction. Parameters of the interactions are taken from
Ref.~\cite{Ita00}.

%Fig.1

\begin{figure}[h]
	\centering
	\includegraphics[width=\figwidth]{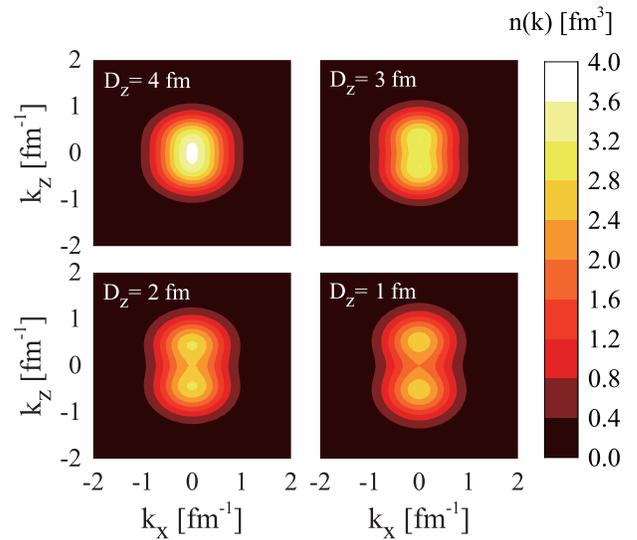}\\
  \caption{The intrinsic nucleon momentum distributions of the
  $\alpha$-$\alpha$ Brink wave function as functions of the $\alpha$-$\alpha$
  distance $D_{z}$ in the $x$-$z$ cross section.}
	\label{fig:brink-xy}
\end{figure}

For the $^{8}$Be nucleus, we calculate the intrinsic momentum distribution
along the $z$-axis described by the $\alpha$-$\alpha$ Brink wave function in
Eq.~(\ref{eq:brink}). This is a toy model with the relative motion of two
$\alpha$-clusters localized around the relative distance
$\bm{D}=\bm{R}_{1}-\bm{R}_{2}$. The $z$-axis is set as the symmetry axis, hence
the relative distance of two $\alpha$-clusters is determined by the parameter
$D_z$. We show in Fig.~\ref{fig:brink-curve} and \ref{fig:brink-xy} the
nucleon momentum distributions of $^{8}$Be with $\alpha$-$\alpha$ distances
$D_z=\{4,3,2,1\}$ fm, which correspond to the evolution from the weak to the
strong overlap between the two $\alpha$-clusters. 

In Fig.~\ref{fig:brink-curve}, the solid and dotted curves are used to
illustrate the momentum distribution of two $\alpha$-clusters without
antisymmetrization between them, where the solid curve shows the distribution
of two free $\alpha$-clusters as discussed before, and the dotted one is for
the $\alpha$-$\alpha$ Brink wave function at infinite distance
$(\bm{D}\to\infty)$.  These two curves have the Gaussian shape, but the one of
the Brink wave function has an enhanced tail part and a smaller value at the
zero momentum. This difference arises from the effect of the localization of
the inter-cluster motion in the $\alpha$-$\alpha$ Brink wave function.
Analytically using Eq.~(\ref{eq:amd-momentum2}), we derive the momentum
distribution $n(k)$ of two free $\alpha$-clusters as
\begin{equation}\label{eq:free-2-alpha}
n^{\textrm{Free}}_{2\alpha}(\boldsymbol{k})
  = \left(\frac{8}{3 \pi \nu}\right)^{3 / 2}
  \textup{exp}\left(-\frac{2}{3 \nu} \boldsymbol{k}^2\right), 
\end{equation}
where $\epsilon=3/4$ for both $\alpha$-clusters. For the $\alpha$-$\alpha$
Brink wave function at infinite distance $D\to\infty$, we obtain
\begin{equation}\label{eq:brink-2-alpha}
n^{\textrm{Infinite-Brink}}_{2\alpha}(\boldsymbol{k})
  = \left(\frac{16}{7 \pi \nu}\right)^{3 / 2}
  \textup{exp}\left(-\frac{4}{7 \nu} \boldsymbol{k}^2\right),
\end{equation}
where $\epsilon=7/8$. The coefficient in Eq.~(\ref{eq:brink-2-alpha}) is
smaller than the one in Eq.~(\ref{eq:free-2-alpha}), which results in a
smaller value at the center for the dotted curve as compared to the solid
curve shown in Fig.~\ref{fig:brink-curve}.

For the curve with finite $D_z=4$ fm in Fig.~\ref{fig:brink-curve}, a Gaussian
shape, which is similar to the dotted curve with infinite $D_{z}$, is observed
but with a slight depression near $k=0$ fm$^{-1}$. This could be explained by
the small overlap between two $\alpha$-clusters and the nucleons are excited
to the relatively higher momentum because of the inter-cluster
antisymmetrization. By further reducing the $\alpha$-$\alpha$ distance $D_z$,
it is clearly seen that the momentum distribution becomes more peripheral and
a dip structure appears at zero-momentum, which corresponds to a rather
compact system where the nucleon excitation is most significant. This process,
in which the Gaussian form of momentum distribution is broken, shows the
dissolution of two $\alpha$-clusters when they are in large spatial overlap
with each other.

%Fig.2

The nucleon dynamics within the overlapped $\alpha$-clusters can be more
clearly demonstrated by the intrinsic density distribution in the momentum
space. The corresponding density values in the $x$-$z$ cross section are shown
in Fig.~\ref{fig:brink-xy} for different $\alpha$-$\alpha$ distances $D_{z}$.
With large distance $D_{z}=4$ fm, the momentum distribution of two
$\alpha$-clusters is almost spherical, which is similar to the Gaussian
distribution predicted for the free $\alpha$-cluster. As the the inter-cluster
distance is decreased, the spherical symmetry is found to be broken and a
large deformation of momentum distribution emerges. The most intriguing
observation is that in the most compact configuration with $D_{z}=1$ fm, a
``cluster structure'' is observed in the intrinsic momentum distribution of
two $\alpha$-clusters, which is astonishing when considering the fact that the
$\alpha$-clusters are strongly dissolved in the coordinate space under this
short relative distance.

\subsection{Nucleon momentum distribution of Beryllium isotopes}
\label{subsec:res-thsr}
We calculate the nucleon momentum distribution for the $^{8-10}$Be isotopes by
using the corresponding THSR wave function for each nucleus, as formulated in
Sec.~\ref{subsec:wave functions}. In our previous works \cite{fun02, Lyu15,
Lyu16}, the accuracy of the THSR wave functions has been proved by reproducing
the physical properties of the $^{8-10}$Be isotopes, such as the energy
spectra and radii. The Hamiltonian is adopted from Ref.~\cite{Lyu15} and the
THSR wave function is variationally determined for each nucleus. The
$\bm{\beta}$ parameters in the optimized THSR wave functions are listed in
Table \ref{table:para}. The resonant state of the $^{8}$Be nucleus is
simulated by a weakly bounded solution that corresponds to a local energy
minimum in the variation of THSR wave function \cite{fun02}.

\begin{table}[h]
	\begin{center}
    \caption{Components of the deformation parameter $\bm{\beta}$ in the
     optimized THSR wave functions of $^{8-10}$Be isotopes. All units are in
     fm.}
    \label{table:para}
    % \begin{tabular}{l c c c l}
    \begin{tabular*}{6.5cm}{ @{\extracolsep{\fill}} l c c c l }
			\hline
      \hline
			& ${\beta}_{\alpha,xy} $ & ${\beta}_{\alpha,z} $ & ${\beta}_{n,xy}$ &
			${\beta}_{n,z}$\\
			\hline
			$^8$Be & 1.0 & 11.0  & - & - \\
			\hline
			$^9$Be & 0.1 & 4.2 & 2.5 & 2.8 \\
			\hline
			$^{10}$Be & 0.1 & 2.5 & 1.9 & 2.9 \\
			\hline
		\end{tabular*}
	\end{center}
\end{table}

%Fig.3
\begin{figure}[h]
	\centering
	\includegraphics[width=\figwidth]{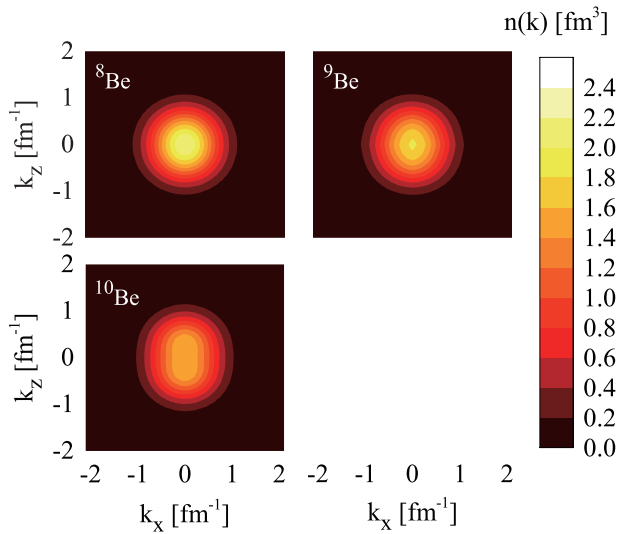}\\
  \caption{The intrinsic proton momentum distributions of $^{8-10}$Be isotopes
   in the $x$-$z$ cross section.}
	\label{fig:thsr-intrinsic}
\end{figure}

From Table \ref{table:para}, it is clearly shown that the $\bm{\beta}$
parameter shrinks when introducing additional valence neutrons into the
nucleus, which corresponds to an evolution from the dilute cluster gas in
$^{8}$Be to the compact structure in $^{10}$Be. The motion of
$\alpha$-clusters in the Be isotopes is demonstrated by the proton momentum
distribution of $^{8-10}$Be isotopes, as shown in
Fig.~\ref{fig:thsr-intrinsic}. For the dilute $^{8}$Be, we obtain spherical
Gaussian distribution for the protons while clear deformation is observed for
the compact $^{10}$Be nucleus, which is similar to the Brink model discussed
in Fig.~\ref{fig:brink-xy}. We note that the deformation in the THSR wave
function of $^{8}$Be is weaker than the Brink wave functions in
Sec.~\ref{subsec:brink}, which is due to the spatially extended motion of
$\alpha$-clusters in $^{8}$Be described by the THSR wave functions.

%Fig.4
\begin{figure}[h]
	\centering
	\includegraphics[width=0.5\textwidth]{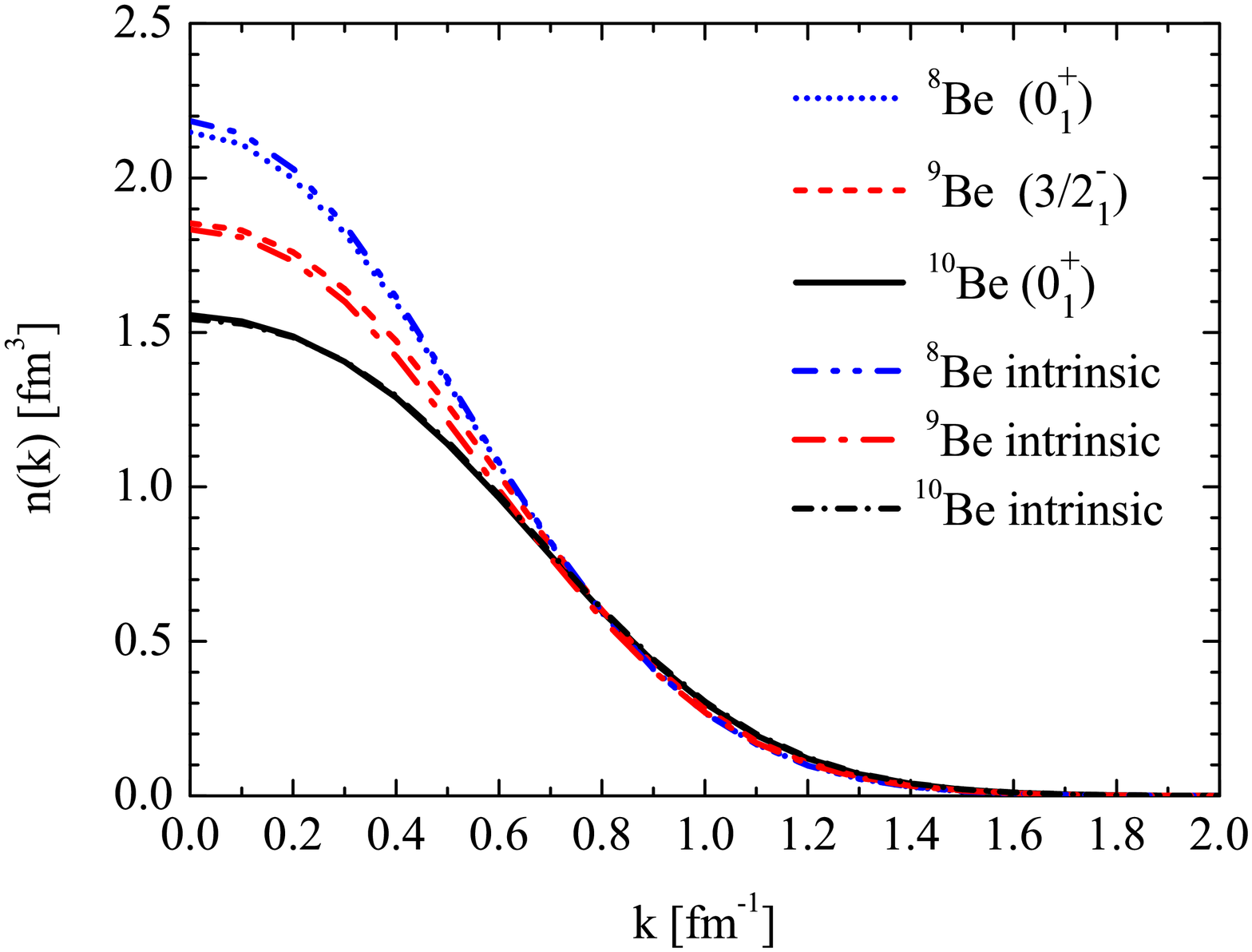}\\
  \caption{The nucleon momentum distributions of the two $\alpha$-clusters
  within the $^{8-10}$Be isotopes. Curves denoted with parentheses are for the
  $^8$Be ($0_{1}^+$), $^9$Be ($3/2_{1}^-$), and $^{10}$Be ($0_{1}^+$) nuclei
  in the laboratory frames. Curves denoted with ``intrinsic" are for the
  intrinsic frames before angular momentum projections. }
	\label{fig:proton-curve}
\end{figure}

To obtain the distribution of two $\alpha$-clusters for the $^{8-10}$Be
isotopes in the laboratory frame, we calculate the angle-averaged momentum
distribution on the sphere surface $S$ with the radius $k$, which is defined
as
\begin{equation}
  n(k)=\frac{1}{4\pi k^{2}} 
        \int_{|\boldsymbol{k}|=k} n(\boldsymbol{k}) {\rm d} S.
\end{equation}
The THSR wave functions are adopted for the $^8$Be$(0_{1}^+)$, $^9$Be$(3/2_{1}^-)$,
and $^{10}$Be$(0_{1}^+)$ isotopes after angular momentum projection
\cite{horiuchi86}, and the calculated distributions are shown in
Fig.~\ref{fig:proton-curve}. In addition, corresponding curves for the
intrinsic frames are also included for comparison, where only slight
differences before and after angular momentum projection are observed. 

For the compact cluster state in $^{10}$Be, the depression at zero-momentum
and the enhanced tail region are observed once again in
Fig.~\ref{fig:proton-curve}, as compared to the Gaussian-like curve of
$^{8}$Be. The depression in the distribution of the two $\alpha$-clusters in
$^{10}$Be is not strong enough to produce a dip structure, but the large
deviation from the $^{8}$Be curve shows clearly the strong antisymmetrization
between $\alpha$-clusters. In $^{9}$Be, relatively weaker $\alpha$-$\alpha$
overlap is found from the intermediate zero-momentum depression. 

%Fig.5
\begin{figure}[htbp]
	\centering
	\includegraphics[width=0.5\textwidth]{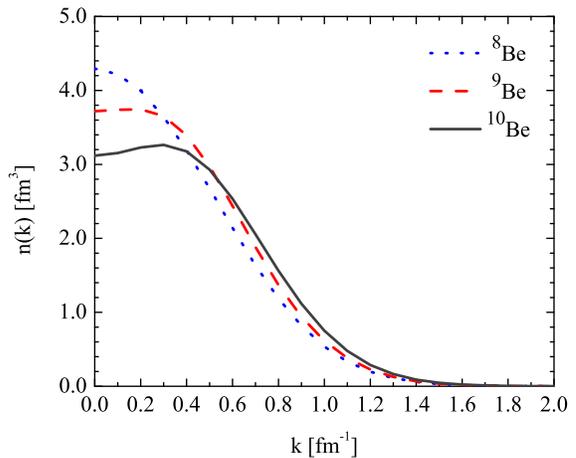}\\
	\caption{The total momentum distributions of the $^8$Be$(0_{1}^+)$,
  $^9$Be$(3/2_{1}^-)$, and $^{10}$Be$(0_{1}^+)$ isotopes with THSR wave
  function.}
	\label{fig:nucleon-curve}
\end{figure}

\begin{figure}[htbp]
	\centering
	\includegraphics[width=0.5\textwidth]{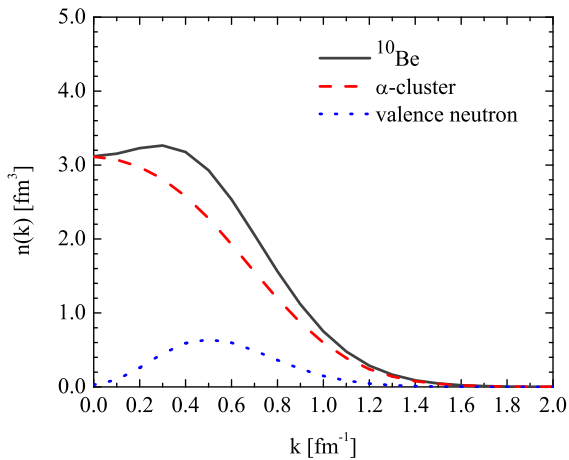}\\
  \caption{Decomposition of momentum distributions in $^{10}$Be ($0_{1}^{+}$)
  into components contributed by $\alpha$-clusters and valence neutrons.}
	\label{fig:components}
\end{figure}

We also calculate the momentum distribution of all nucleons in each isotope,
as shown in Fig.~\ref{fig:nucleon-curve}. The solid curve for the $^{8}$Be
nucleus is not changed from Fig.~\ref{fig:proton-curve}, except the different
amplitude. However, the dip structures appear in the curves of $^{9}$Be and
$^{10}$Be isotopes, with further enhanced tail region, as compared to
Fig.~\ref{fig:proton-curve}. In Ref.~\cite{ciofi96}, similar dip structures
have been observed in experimental results for $^{12}$C and $^{16}$O nuclei.
We note that the dip structures in Fig.~\ref{fig:nucleon-curve} are also
contributed by the $p$-shell occupation by valence neutrons, in addition to
the $\alpha$-$\alpha$ antisymmetrization which is the major origin of dip
structure in Fig.~\ref{fig:brink-curve}. This conjecture is proved by the
decomposition of the momentum distribution in $^{10}$Be, as shown in
Fig.~\ref{fig:components}. Here, the momentum component contributed by the
$\alpha$-clusters and the valence neutrons are calculated by replacing the
summation over $i$ in Eq.~(\ref{eq:amd-momentum}) with the corresponding set
of nucleon indices. In this figure, the contribution from valence neutrons is
presented by the blue dotted curve, in which the node structure of $p$-wave is
clearly observed. It is also found that the valence neutrons contribute mostly
around $k\approx0.5$ fm$^{-1}$, which enhances the tail region in
Fig.~\ref{fig:nucleon-curve}. We note that using the relation $\braket{k}=2\nu
D$ in Ref.~\cite{myo17e}, this $k$ corresponds to the high-momentum excitation
of neutron with imaginary shift of about $D=1$ fm in coordinate space, which
is the mean location of valence neutron measured from the center of mass.

%Fig.6

\section{Conclusion}
\label{sec:conclusion}
We investigated the evolution of clustering structure through the momentum
distributions in the $^{8-10}$Be isotopes, which are calculated by using the
analytical expressions formulated in our recent work. The general features of
the nucleon dynamics within two $\alpha$-cluster system under
antisymmetrization have been discussed via the momentum distribution of a
Brink type $\alpha$-$\alpha$ wave function. For the state with a strong
inter-cluster overlap at small relative distance, we observed a significant
depression with a dip structure at zero-momentum and an enhanced tail at
relatively higher momentum region, which is a clear manifestation for the
nucleon momentum excitation induced by the antisymmetrization between two
$\alpha$-clusters. The most interesting observation is that the momentum
distribution of the extremely compact $\alpha$-$\alpha$ system shows a
``cluster structure'' in the intrinsic frame of momentum space, which is
complementary to its significant $\alpha$-cluster dissolution in the
coordinate space because of the strong antisymmetrization. 

For the physical nuclei, we adopted the THSR wave functions for the clustering
states in the $^{8-10}$Be isotopes, which provided successful descriptions for
these nuclei in our previous works. The evolution from the dilute gas-like
state to the compact one with $\alpha$-cluster dissolution was demonstrated by
the calculated nucleon distribution for the two $\alpha$-clusters within the
Be isotopes, where the successive depression at zero-momentum was observed in
the curves of $^{9-10}$Be, as compared to the curve of $^{8}$Be. We also
calculated the momentum distribution of total nucleons for the $^{8-10}$Be
isotopes, and observe a significant depression at zero-momentum for the
compact $^{10}$Be nucleus. We performed the decomposition for the momentum
distribution and found that both the inter-cluster antisymmetrization and the
$p$-orbit occupation by the valence neutrons contribute to the dip structure
at zero-momentum in the $^{10}$Be nucleus. In this study, we propose a new
window for the investigations of $\alpha$-clustering structures in the
$^{8-10}$Be isotopes by connecting the lower region of momentum distributions
with the nuclear clustering structures, which could be extended to the future
studies of heavier nuclear clustering states.

% \newpage \section{Appendix:} \subsection{Parameters of the THSR wave
% functions}
\section*{Acknowledgments}
The authors would like to thank Prof.~Hiroki Takemoto, Dr.~Niu Wan,
Prof.~Masaaki Kimura and Prof.~Bo Zhou for the valuable discussions, and also
the anonymous reviewer for the valuable comments. This work is supported by
the National Key Research and Development Program of China (Grants No.
2018YFA0404403 and No. 2016YFE0129300), the National Natural Science
Foundation of China (Grants No. 11975167, No. 11961141003, No. 11761161001,
No. 11535004, No. 11881240623, No. 11822503 and No. 11575082), the Science and
Technology Development Fund of Macau under Grant No. 008/2017/AFJ, and the
JSPS KAKENHI Grants No. JP18K03660. The author M.L. acknowledges the support
from the RCNP theoretical group for his stay in RCNP and the fruitful
discussions with the members, and the support from the Yozo Nogami Research
Encouragement Funding. The author Q.Z. is grateful to the members of the
nuclear theory group in Hokkaido University for fruitful discussions.

% \subsection*{References}
% \vfill\pagebreak
%%%%%%%%%%%%%%%%%%%%%%%%%%%%%%%%%%%%%%%%%%%%%%%%%%%%%%%%%%%%%
%%%%%%%%%%%%%%%%%%%%%%%%%%%%%%%%%%%%%%%%%%%%%%%%%%%%%%%%%%%%%
\def\JL#1#2#3#4{ {{\rm #1}} \textbf{#2}, #3 (#4)}  % Physical Review
\nc{\PR}[3]     {\JL{Phys. Rev.}{#1}{#2}{#3}}
\nc{\PRC}[3]    {\JL{Phys. Rev.~C}{#1}{#2}{#3}}
\nc{\RMP}[3]    {\JL{Rev. Mod. Phys.}{#1}{#2}{#3}}
\nc{\PRA}[3]    {\JL{Phys. Rev.~A}{#1}{#2}{#3}}
\nc{\PRL}[3]    {\JL{Phys. Rev. Lett.}{#1}{#2}{#3}}
\nc{\NP}[3]     {\JL{Nucl. Phys.}{#1}{#2}{#3}}
\nc{\NPA}[3]    {\JL{Nucl. Phys.}{A#1}{#2}{#3}}
\nc{\PL}[3]     {\JL{Phys. Lett.}{#1}{#2}{#3}}
\nc{\PLB}[3]    {\JL{Phys. Lett.~B}{#1}{#2}{#3}}
\nc{\PTP}[3]    {\JL{Prog. Theor. Phys.}{#1}{#2}{#3}}
\nc{\PTPS}[3]   {\JL{Prog. Theor. Phys. Suppl.}{#1}{#2}{#3}}
\nc{\PTEP}[3]   {\JL{Prog. Theor. Exp. Phys.}{#1}{#2}{#3}}
\nc{\PRep}[3]   {\JL{Phys. Rep.}{#1}{#2}{#3}}
\nc{\PPNP}[3]   {\JL{Prog.\ Part.\ Nucl.\ Phys.}{#1}{#2}{#3}}
\nc{\JPG}[3]    {\JL{J. of Phys. G}{#1}{#2}{#3}}
\nc{\andvol}[3] {{\it ibid.}\JL{}{#1}{#2}{#3}}
%%%%%%%%%%%%%%%%%%%%%%%%%%%%%%%%%%%%%%%%%%%%%%%%%%%%%%%%%%%%%

\end{document}